\documentclass[12pt]{article}
\usepackage{amssymb,amsmath,epsfig}

\begin{document}

\title{\bf Newtonian and Post Newtonian Expansionfree Fluid
Evolution in $f(R)$ Gravity}

\author{M. Sharif \thanks{msharif.math@pu.edu.pk} and H. Rizwana
Kausar\thanks{rizwa\_math@yahoo.com}\\
Department of Mathematics, University of the Punjab,\\
Quaid-e-Azam Campus, Lahore-54590, Pakistan.}

\date{}
\maketitle

\begin{abstract}
We consider a collapsing sphere and discuss its evolution under
the vanishing expansion scalar in the framework of $f(R)$ gravity.
The fluid is assumed to be locally anisotropic which evolves
adiabatically. To study the dynamics of the collapsing fluid,
Newtonian and post Newtonian regimes are taken into account. The
field equations are investigated for a well-known $f(R)$ model of
the form $R+\delta R^2$ admitting Schwarzschild solution. The
perturbation scheme is used on the dynamical equations to explore
the instability conditions of expansionfree fluid evolution. We
conclude that instability conditions depend upon pressure
anisotropy, energy density and some constraints arising from this
theory.
\end{abstract}

\section{Introduction}

To accommodate observational data with theoretical predictions, one
needs to introduce the dark energy (DE) contributions in the most
successful gravitational theory of General Relativity (GR). On the
other hand, modified theories of gravity may provide a cosmological
accelerating mechanism without introducing any extra DE
contribution. Modifications of GR by $f(R)$ models are able to mimic
the standard $\Lambda$CDM cosmological evolution. These models may
have vacuum solutions with null scalar curvature that allow to
recover some GR solutions. These may also lead to the existence of
some new solutions particularly in spherically symmetric scenario.

In the last decade, some $f(R)$ models were considered to modify
GR at small scales to explain inflation, e.g., $f(R)\propto R^2$
but failed to explain the late time acceleration. A model like
$f(R)\propto\frac{1}{R}$ was proposed to explain this acceleration
but attained no interest due to conflict with solar system tests
(Chiba 2003; Dolgov and Kawasaki 2003). A cosmological viable
model needs to satisfy the evolution of big-bang nucleosynthesis,
radiation and matter dominated eras. Also, they must provide
cosmological perturbations compatible with cosmological
constraints from cosmic microwave background and large scale
structure. The problem of cosmological perturbations in this
modified theory and their consequences have widely been discussed
in literature, e.g., (Hwang and Noh 2006; Carroll et al. 2006;
Carloni et al. 2008; Tsujikawa et al. 2008). Since the lagrangian
$R+f(R)$ is analytic at $R=0$, so the Schwarzschild and other
important GR solutions (without cosmological constant) are also
the solutions of $f(R)$ gravity (de la Cruz-Dombrize and Dobado
2006).

It is known that $f(R)$ models reveal black hole solutions as in GR,
therefore it is quite natural to discuss the question of black hole
features and dynamics of the gravitational collapse in this modified
theory. When we take conformal transformation of $f(R)$ action, it
is found that Schwarzschild solution is the only static spherically
symmetric solution for a model of the form $R+\delta R^2$ (Whitt
1984). Also, the uniqueness theorems of spherically symmetric
solutions for general polynomial action in arbitrary dimension were
proposed by using conformal transformation (Mignemi and Wiltshire
1992). Work on black hole solutions in $f(R)$ gravity has been
carried out by many authors, e.g., (Moon et al. 2011; de la
Cruz-Dombriz et al. 2009; Mazharimousavi et al. arXiv/1105.3659v3)

It is found that evolution of expansionfree spherically symmetric
distribution is consistent with the existence of a vacuum cavity
within the matter distribution (Herrera and Santos 2004; Herrera
et al. 2008 ). Skripkin model is the first example of
expansionfree evolution (Skripkin 1960). This model corresponds to
evolution of spherically symmetric perfect fluid distribution with
constant energy density. We have explored this model in $f(R)$
gravity (Sharif and Kausar 2011) and effects of $f(R)$ DE on the
dynamics of dissipative fluid collapse (Sharif and Kausar 2010;
Sharif and Kausar 2011). Also, solution of the field equations for
Bianchi models and spherically symmetric dust solution are
obtained in this theory (Sharif and Kausar 2011; Sharif and Kausar
2011; Sharif and Kausar 2011).

The physical applications of such models lie at the core of
astrophysical background where a cavity within the fluid
distribution is present. It may help to study the formation of
voids on the cosmological scales (Liddle and Wands 1991). Also,
the expansionfree fluid evolution causes a blowup of shear scalar
which results the appearance of a naked singularity (Joshi et al.
2002). It is interesting to note that expansionfree models defines
the two hypersurfaces, one separating the fluid distribution
externally from the Schwarzschild vacuum solution while other one
is the boundary between the central Minkowskian cavity and the
fluid. Furthermore, expansionfree models require anisotropy in
pressure and inhomogeneity in energy density.

The existence of stellar model can be assured if it is stable
against fluctuations. The problem of dynamical instability is
closely related to the structure formation and evolution of
self-gravitating objects. A pioneer work in this direction was done
by Chandrasekhar (Chandrasekhar 1964). Afterwards, this issue has
been investigated by many authors for adiabatic, non-adiabatic,
anisotropic and shearing viscous fluids (Herrera et al. 1989; Chan
et al. 1989; Chan et al. 1994; Chan et al 1993; Herrera et al,
arXiv/1010.1518). The dynamical instability of collapsing fluids can
be well discussed in term of adiabatic index $\Gamma$. It measures
the variation of pressure with respect to a given variation of
energy density and defines the range of instability. In GR, the
dynamical instability is independent of this $\Gamma$ factor
(Herrera et al, arXiv/1010.1518) in the Newtonian and post Newtonian
regimes. In a recent paper (Sharif and Kausar 2011), we have
investigated the dynamical instability of expansionfree fluid
evolution in $f(R)$ gravity by assuming all the higher order
curvature terms on the matter side. This provides the effects of
$f(R)$ DE on the instability conditions.

In this paper, we treat the field equation in the usual fourth order
form and discuss instability conditions of expansionfree fluid
collapse in metric $f(R)$ theory. Here we use Ricci scalar to
evaluate time dependent part of the perturbed quantities. We
consider locally anisotropic fluid distribution inside the
collapsing sphere. The format of the paper is as follows. In section
\textbf{2}, we formulate the field equations and dynamical equations
for spherically symmetric spacetime in $f(R)$ gravity by taking
locally anisotropy in the pressure. Section \textbf{3} is devoted to
study the perturbation scheme on physical quantities and the metric
coefficients by assuming that fluid initially is in static
equilibrium. A well-known physical $f(R)$ model is considered and
perturbed on the same pattern. In section \textbf{4}, Newtonian and
post Newtonian regimes are taken into account. Discussion is made on
the dynamical instability conditions of the expansionfree fluid
evolution in the Newtonian approximation. The last section
\textbf{5} concludes the main results of the paper.

\section{Field Equations and Dynamical Equations}

We take spherically symmetric distribution of anisotropic
collapsing fluid in co-moving coordinates. The line element is
given by
\begin{equation}\setcounter{equation}{1}\label{6}
ds^2=A^2(t,r)dt^{2}-B^2(t,r)dr^{2}-C^2(t,r)(d\theta^{2}+\sin^2\theta
d\phi^{2}).
\end{equation}
This yields two radii: one is the areal radius $C(r,t)$ measuring
the radius from spherical surface while other is the proper radius
found from $\int B(t,r)dr$. Also, there is a hypersurface
${\Sigma^{(e)}}$ separating interior spacetime from the exterior
one. The energy-momentum tensor for locally anisotropic fluid can be
written as
\begin{equation}\label{8}
T_{\alpha\beta}=(\rho+p_{\perp})u_{\alpha}u_{\beta}-p_{\perp}g_{\alpha\beta}
+(p_r-p_{\perp})\chi_{\alpha}
\chi_{\beta},\quad(\alpha,\beta=0,1,2,3),
\end{equation}
where $\rho$ is the energy density, $p_{\perp}$ the tangential
pressure, $p_r$ the radial pressure, $u_{\alpha}$ the
four-velocity of the fluid and $\chi_{\alpha}$ is the unit
four-vector along the radial direction. These quantities satisfy
the relations
\begin{equation}\label{9}
u^{\alpha}u_{\alpha}=1,\quad\chi^{\alpha}\chi_{\alpha}=-1,\quad
\chi^{\alpha}u_{\alpha}=0
\end{equation}
which are obtained from the following definitions in co-moving
coordinates
\begin{equation}\label{10}
u^{\alpha}=A^{-1}\delta^{\alpha}_{0},\quad
\chi^{\alpha}=B^{-1}\delta^{\alpha}_{1}.
\end{equation}
The acceleration $a_{\alpha}$ and the expansion $\Theta$ are given
by
\begin{equation}\label{12}
a_{\alpha}=u_{\alpha;\beta}u^{\beta},\quad
\Theta=u^{\alpha}_{;\alpha}.
\end{equation}
Using Eqs.(\ref{10}) and (\ref{12}), it follows that
\begin{equation}\label{15}
a_{1}=-\frac{A'}{A},\quad a^2=a^\alpha
a_\alpha=\left(\frac{A'}{AB}\right)^2,\quad
\Theta=\frac{1}{A}\left(\frac{\dot{B}}{B}+2\frac{\dot{C}}{C
}\right),
\end{equation}
where dot and prime represent derivatives with respect to $t$ and
$r$ respectively.

The Einstein-Hilbert action in GR is
\begin{equation}\label{1}
S_{EH}=\frac{1}{2 \kappa}\int d^{4}x\sqrt{-g}R,
\end{equation}
where $\kappa$ is the coupling constant, $R$ is the Ricci scalar and
$g$ is the metric tensor. For $f(R)$ theory of gravity, it is
modified as follows
\begin{equation}\label{2}
S_{modif}=\frac{1}{2\kappa}\int d^{4}x\sqrt{-g}f(R),
\end{equation}
where $f(R)$ is the general function of the Ricci scalar. Varying
this action with respect to the metric tensor, we obtain the
following field equations
\begin{equation}\label{3}
F(R)R_{\alpha\beta}-\frac{1}{2}f(R)g_{\alpha\beta}-\nabla_{\alpha}
\nabla_{\beta}F(R)+ g_{\alpha\beta} \Box F(R)=\kappa
T_{\alpha\beta},
\end{equation}
where $F(R)\equiv df(R)/dR$. For spherically symmetric spacetime,
these equations become
\begin{eqnarray}\nonumber
&&\frac{AA''}{B^2}-\frac{\ddot{B}}{B}+\frac{\dot{A}\dot{B}}{AB}-\frac{AA'B'}{B^3}
-\frac{2\ddot{C}}{C}
+\frac{2\dot{A}\dot{C}}{AC}+\frac{2AA'C'}{B^2C}-\frac{A^2}{2}\frac{f(R)}{F}\\\label{16}
&-&\frac{A^2{F''}}{B^2F}-\frac{\dot{F}}{F}\left(\frac{-\dot{B}}{B}+\frac{2\dot{C}}{C}\right)
-\frac{F'A^2}{F B^2}\left(\frac{-B'}{B}+\frac{2C'}{C}\right)
=\kappa{\rho}A^{2},\\\label{17}
&-&2\left(\frac{\dot{C'}}{C}-\frac{\dot{C}A'}{CA}-\frac{\dot{B}C'}{BC}\right)-\frac{\dot{F'}}{
F}+\frac{A'\dot{F}}{AF}+\frac{\dot{B}F'}{BF}=0,\\\nonumber
&-&\frac{A''}{A}+\frac{B\ddot{B}}{A^2}-\frac{\dot{A}\dot{B}B}{A^3}+\frac{A'B'}{AB}-
\frac{2C''}{C}
+\frac{2\dot{C}\dot{B}B}{A^2C}+\frac{2B'C'}{BC}+\frac{B^2}{2}\frac{f(R)}{F}\\\label{18}
&-&\frac{B^2\ddot{F}}{A^2F}+\frac{\dot{F}B^2}{FA^2}\left(\frac{\dot{A}}{A}
+\frac{2\dot{C}}{C}\right)
+\frac{F'}{F}\left(\frac{A'}{A}+\frac{2C'}{C}\right)=\kappa
p_rB^{2},\\\nonumber
&&\frac{\ddot{C}}{CA^2}-\frac{\dot{A}\dot{C}}{CA^3}-\frac{A'C'}{B^2AC}-\frac{C''}{B^2C}
+\frac{\dot{C}\dot{B}}{A^2BC}
+\frac{B'C'}{B^3C}+\frac{1}{C^2}\\\nonumber
&+&\frac{\dot{C}^2}{A^2C^2}-\frac{C'^2}{B^2C^2}
+\frac{1}{2}\frac{f(R)}{F}-\frac{\ddot{F}}{FA^2}+\frac{F''}{FB^2}
+\frac{\dot{F}}{A^2}\left(\frac{\dot{A}}{A}-\frac{\dot{B}}{B}+\frac{\dot{C}}{C}\right)
\\\label{19}
&+&\frac{F'}{B^2}\left(\frac{A'}{A}-\frac{B'}{B}+\frac{C'}{C}\right)
=\kappa p_{\perp}.
\end{eqnarray}
The Ricci scalar curvature is given by
\begin{eqnarray}\nonumber
R&=&2\left[\frac{A''}{AB^2}-\frac{\ddot{B}}{A^2B}+\frac{\dot{A}\dot{B}}{A^3B}
-\frac{A'B'}{AB^3}+\frac{2\ddot{C}}{CA^2}+\frac{2\dot{A}\dot{C}}{ACB^2}\right.\\\label{R}
&+&\left.\frac{2C''}{CB^2}-\frac{2\dot{C}\dot{B}}{CA^2B}
-\frac{2C'B'}{CB^3}-\frac{1}{C^2}-\frac{\dot{C}^2}{A^2C^2}+\frac{C'^2}{B^2C^2}\right].
\end{eqnarray}
For the exterior spacetime to $\Sigma^{(e)}$, we take the
Schwarzschild spacetime in the form
\begin{equation}\label{7}
ds^2=\left(1-\frac{2M}{r}\right)d\nu^2+2drd\nu-r^2(d\theta^2+\sin^2\theta
d\phi^2),
\end{equation}
where $M$ represents the total mass and $\nu$ is the retarded
time.

In order to study the properties of collapsing process, we
formulate the dynamical equations. For this purpose, we use the
Misner-Sharp mass function defined as follows (Misner and Sharp
1964)
\begin{equation}\label{20}
m(t,r)=\frac{C}{2}(1+g^{\mu\nu}C_{,\mu}C_{,\nu})=\frac{C}{2}\left(1+\frac{\dot{C}^2}{A^2}
-\frac{C'^2}{B^2}\right).
\end{equation}
This equation provides the total energy inside a spherical body of
radius "C". From the continuity of the first and second
fundamental forms, the matching of the adiabatic sphere to the
Schwarzschild spacetime on the boundary surface, ${\Sigma^{(e)}}$,
yields
\begin{equation}\label{j1}
M\overset{\Sigma^{(e)}}{=}m(t,r).
\end{equation}
The proper time and radial derivatives are given by
\begin{equation}\label{45}
D_{T}=\frac{1}{A}\frac{\partial}{\partial t},\quad
D_{C}=\frac{1}{C'}\frac{\partial}{\partial r},
\end{equation}
where $C$ is the areal radius of the spherical surface. The velocity
of the collapsing fluid is defined as
\begin{equation}\label{22}
U=D_{T}C=\frac{\dot{C}}{A}
\end{equation}
which is always negative. Using this expression, Eq.(\ref{20})
implies that
\begin{equation}\label{23}
E\equiv\frac{C'}{B}=\left[1+U^{2}+\frac{2m}{C}\right]^{1/2}.
\end{equation}
The dynamical equations can be obtained from the non-trivial
contracted components of the Bianchi identities. Consider the
following two equations
\begin{eqnarray}\label{52}
T^{\alpha\beta}_{;\beta}u_{\alpha}=0,\quad T^{\alpha\beta}_{;\beta}
\chi_{\alpha}=0
\end{eqnarray}
which yield
\begin{eqnarray}\label{28}
\dot{\rho}+(\rho+p_r)\frac{\dot{B}}{B}+2(\rho+p_{\perp})
\frac{\dot{C}}{C}&=&0,\\\label{29}
p_r'+(\rho+p_r)\frac{A'}{A}+2(p_r-p_{\perp})\frac{C'}{C}&=&0.
\end{eqnarray}

\section{The $f(R)$ Model and Perturbation Scheme}

In this section, we consider a particular $f(R)$ model and apply
perturbation scheme on all the above equations. Consider the
following well-known $f(R)$ model
\begin{equation}\setcounter{equation}{1}\label{fr}
f(R)=R+\delta R^2,
\end{equation}
where $\delta$ is any positive real number. The stability criteria
for this model corresponds to $f^{''}(R)>0$. For $\delta=0$, GR is
recovered in which black holes are stable classically but not
quantum mechanically due to Hawking radiations. Since such
features are also found in $f(R)$ gravity, hence the classical
stability condition for the Schwarzschild black hole can be
enunciated as $f^{''}(R)>0$ (Sotiriou and Faraoni 2010).

The purpose of introducing perturbation scheme is to analyze the
instability conditions of the dynamical equation. We assume that
initially all the quantities have only radial dependence, i.e.,
fluid is in static equilibrium. Then these quantities have time
dependence as well in their perturbation, i.e.,
\begin{eqnarray}\label{41'}
A(t,r)&=&A_0(r)+\epsilon T(t)a(r),\\\label{42'}
B(t,r)&=&B_0(r)+\epsilon T(t)b(r),\\\label{43'}
C(t,r)&=&C_0(r)+\epsilon T(t)\bar{c}(r),\\\label{44'}
\rho(t,r)&=&\rho_0(r)+\epsilon {\bar{\rho}}(t,r),\\\label{45'}
p_r(t,r)&=&p_{r0}(r)+\epsilon {\bar{p_r}}(t,r),\\\label{46'}
p_{\perp}(t,r)&=&p_{\perp0}(r)+\epsilon{\bar{p_{\perp}}}(t,r),\\\label{47'}
m(t,r)&=&m_0(r)+\epsilon {\bar{m}}(t,r),\\\label{48'}
\Theta(t,r)&=&\epsilon {\bar{\Theta}}(t,r).
\end{eqnarray}
Also, the Ricci scalar in $f(R)$ model leads to
\begin{eqnarray}\label{49'}
R(t,r)&=&R_0(r)+\epsilon T(t)e(r),\\\label{50'} f(R)&=&R_0(1+2\delta
R_0)+\epsilon T(t)e(r)(1+2\delta R_0),\\\label{51'}
F(R)&=&(1+2\delta R_0)+2\epsilon \delta T(t)e(r),
\end{eqnarray}
where $0<\epsilon\ll1$. We choose radial part of the areal radius as
the Schwarzschild coordinate, i.e., $C_0(r)=r$.

Using the above  values in (\ref{16})-(\ref{19}), the static
configuration turns out to be
\begin{eqnarray}\nonumber
&&\frac{A_0''}{A_0}-\frac{A_0'B_0'}{A_0B_0}+\frac{2A_0'}{A_0r}
-\frac{B_0^2}{2}\frac{R_0(1+2\delta R_0)}{1+2\delta
R_0}\\\label{16s} &-&\frac{2\delta R_0''}{1+2\delta
R_0}-\frac{2\delta R_0'}{1+2\delta
R_0}\left(\frac{-B_0'}{B_0}+\frac{2}{r}\right)=\kappa{\rho_0}B_0^{2},\\\nonumber
&-&\frac{A_0''}{A_0}+\frac{A_0'B_0'}{A_0B_0}+\frac{2B_0'}{B_0r}
+\frac{B_0^2}{2}\frac{R_0(1+2\delta R_0)}{1+2\delta
R_0}\\\label{18s} &+&\frac{2\delta R_0'}{1+2\delta
R_0}\left(\frac{A_0'}{A_0}+\frac{2}{r}\right)=\kappa
p_rB_0^{2},\\\nonumber &-&\frac{A_0'}{A_0r}
+\frac{B_0'}{B_0r}+\frac{B_0^2}{r^2}-\frac{1}{r^2}
+\frac{B_0^2}{2}\frac{R_0(1+2\delta R_0)}{1+2\delta
R_0}+\frac{2\delta R_0''}{1+2\delta R_0}
\\\label{19s}
&+&\frac{2\delta R_0'}{1+2\delta
R_0}\left(\frac{A_0'}{A_0}-\frac{B_0'}{B_0}+\frac{1}{r}\right)
=\kappa p_{\perp}B_0^2.
\end{eqnarray}
The static part of the Ricci scalar takes the form
\begin{equation}\label{Rs}
R_0(r)=\frac{2}{B_0^2}\left[\frac{A_0''}{A_0}-\frac{A_0'B_0'}{A_0B_0}
-\frac{2B_0'}{B_0r}-\frac{B_0^2}{r^2}+\frac{1}{r^2}\right].
\end{equation}

Using the perturbed quantities given in Eqs.(\ref{41'})-(\ref{51'})
along with (\ref{16s})-(\ref{19s}) in the field equations, it
follows that
\begin{eqnarray}\nonumber
&&\frac{aA_0''}{A_0}+a''-2bA_0''-\frac{aA_0'B_0'}{A_0B_0}-\frac{a'B_0'}{B_0}
-\frac{b'A_0'}{B_0A_0}+\frac{3bB_0'A_0'}{B_0^2}+\frac{2aA_0'}{A_0r}+\frac{2a'}{r}\\\nonumber
&-&\frac{2A_0'b}{B_0r}-\frac{2A_0'\bar{c}}{r^2}-\frac{2\bar{c}'A_0'}{r}-\frac{aB_0^2(1+\delta
R_0)}{A_0(1+2\delta R_0)}-\frac{eB_0^2}{2A_0}-\frac{e\delta
A_0B_0R_0(1+\delta R_0)}{(1+2\delta R_0)^2}\\\nonumber
&+&\frac{2\delta}{1+\delta
R_0}\left[-\frac{2a}{A_0}+\frac{2bA_0}{B_0}-e''A_0+2\delta
eR_0''A_0B_0^2+\frac{2aB_0'R_0'}{A_0B_0}-\frac{2bA_0B_0'R_0'}{B_0^2}\right.\\\nonumber
&+&\left.\frac{e' A_0'B_0'}{B_0}-\frac{2e\delta A_0
R_0'B_0'}{(1+2\delta R_0)B_0}+\frac{4a R_0'}{rA_0}+\frac{4bA_0
R_0'}{B_0r}+\frac{2e'A_0}{r}+\frac{4e\delta A_0 R_0'}{r(1+2\delta
R_0)}\right.\\\nonumber &+&\left.
A_0R_0'\left(\frac{b}{B_0}\right)'-2A_0R_0'\left(\frac{\bar{c}}{r}\right)'\right]
-\frac{\ddot{T}}{T}\left(\frac{b}{B_0}+\frac{\bar{c}}{r}\right)=\frac{\kappa
\bar{\rho}A_0B_0^{2}}{T}+2\kappa
\rho_0aB_0^2,\\\label{16p}\\\label{17p}
&&\left(\frac{\bar{c}}{r}\right)'-\frac{b}{B_0r}-\frac{\bar{c}A_0'}{rA_0}
-\frac{\delta}{1+2\delta
R_0}\left[-e'+e\frac{A_0'}{A_0} +\frac{bR_0'}{B_0}\right]=0,
\end{eqnarray}
\begin{eqnarray}\nonumber
&&\frac{aA_0''}{A_0^2}-\frac{a''}{A_0}+\frac{a'B_0'}{A_0B_0}+\frac{b'A_0'}{A_0B_0}
+\frac{2b'}{B_0r}+\frac{2\bar{c}'B_0'}{B_0^2r}-\frac{2bB_0'}{B_0r^2}
-\frac{2\bar{c}B_0'}{r^2B_0}-\frac{2\bar{c}''}{r}\\\nonumber &+&
2\delta\left(\frac{e}{1+2\delta
R_0}\right)'\left(\frac{A_0'}{A_0}+\frac{2}{r}\right)+\frac{1}{1+2\delta
R_0}\left[R_0(1+\delta R_0)\left(b+\frac{\delta e B_0^2}{1+2\delta
R_0}\right) \right.\\\nonumber &+&\left.\frac{eB_0^2}{2}(1+2\delta
R_0)+2\delta
R_0'\left\{\left(\frac{a}{A_0}\right)'+2\left(\frac{\bar{c}}{r}\right)'\right\}\right]+
\frac{\ddot{T}}{T}\frac{B_0}{A_0^2}\left(b-\frac{2e\delta
B_0}{1+2\delta R_0}\right)\\\label{18p}
&=&\frac{\kappa\bar{p_r}B_0^{2}}{T}+\kappa bB_0 p_{r0}, \\\nonumber
&&\frac{1}{r}\left\{\left(\frac{a}{A_0}\right)'+\left(\frac{b}{B_0}\right)'\right\}
+\left(\frac{\bar{c}}{r}\right)'\left\{\frac{A_0'}{A_0}+\frac{B_0'}{B_0}\right\}
+\left(\frac{\bar{c}}{r}\right)''+\frac{2}{r^2}\left(\frac{b}{B_0}+
\frac{\bar{c}}{r}\right)\\\nonumber
&+&\frac{2B_0^2}{r^2}\left(\frac{b}{B_0}-\frac{\bar{c}}{r}\right)
-\frac{aA_0'}{rA_0}-\frac{4b}{B_0r}
\left(\frac{A_0'}{A_0}+\frac{B_0'}{B_0}\right)-\frac{bB_0R_0(1+\delta
R_0)}{1+2\delta R_0}\\\nonumber &+&\frac{2\delta R_0'}{1+2\delta
R_0}\left\{\left(\frac{a}{A_0}\right)'-\left(\frac{b}{B_0}\right)'
+\left(\frac{\bar{c}}{r}\right)'
\right\}-\frac{eB_0^2}{2}-\frac{e\delta B_0^2R_0(1+\delta
R_0)}{(1+2\delta R_0)^2}\\\nonumber &+&\frac{2\delta}{1+2\delta
R_0}\left(e''-\frac{2\delta eR_0''}{1+2\delta R_0}\right)+
2\delta\left(\frac{e}{1+2\delta
R_0}\right)'\left(\frac{A_0'}{A_0}-\frac{B_0'}{B_0}+\frac{2}{r}\right)\\\label{19p}
&-&\frac{\ddot{T}}{T}\frac{B_0^2}{A_0^2}\left(\frac{\bar{c}}{r}+\frac{2e\delta}{1+2\delta
R_0} \right)=\frac{\kappa\bar{p}_{\perp}B_0^{2}}{T}+2\kappa bB_0
p_{\perp0}.
\end{eqnarray}
We can write the perturbed configuration of the Ricci scalar
curvature, by using Eq.(\ref{Rs}), as follows
\begin{eqnarray}\nonumber
&-&\frac{1}{A_0B_0}\left(a''-\frac{aA_0''}{A_0}-\frac{2bA_0''}{B_0}\right)
-\frac{2}{rB_0^3}\left(b'+\bar{c}'B_0'-\frac{\bar{c}B_0'}{r}-
\frac{3bB_0'}{B_0}\right)\\\nonumber
&+&\frac{e}{2}+\frac{\bar{c}}{r^3}-\frac{1}{A_0B_0^3}\left(a'B_0'+bA_0'-\frac{aA_0'B_0'}{A_0}
-\frac{3bA_0'B_0'}{B_0}\right)\\\label{Rp}
&+&\frac{2\bar{c}''}{rB_0^2}+\frac{1}{B_0^2r^2}\left(\bar{c}'-\frac{b}{B_0}
-\frac{\bar{c}}{r}\right)-\frac{\ddot{T}}{T}\left(\frac{b}{B_0}
-\frac{2\bar{c}}{r}\right)\frac{1}{A_0^2}=0.
\end{eqnarray}
This equation can also be written as
\begin{equation}\label{66}
\ddot{T}(t)-\alpha(r) T(t)=0,
\end{equation}
where
\begin{eqnarray}\nonumber
\alpha(r)&=&\left[-\frac{1}{A_0B_0}\left(a''-\frac{aA_0''}{A_0}-\frac{2bA_0''}{B_0}\right)
-\frac{2}{rB_0^3}\left(b'+\bar{c}'B_0'-\frac{\bar{c}B_0'}{r}-
\frac{3bB_0'}{B_0}\right)\right.\\\nonumber
&+&\left.\frac{e}{2}+\frac{\bar{c}}{r^3}-\frac{1}{A_0B_0^3}\left(a'B_0'+bA_0'
-\frac{aA_0'B_0'}{A_0}
-\frac{3bA_0'B_0'}{B_0}\right)\right.\\\label{67}
&+&\frac{1}{B_0^2r^2}\left(\bar{c}'-\frac{b}{B_0}
-\frac{\bar{c}}{r}\right)+\left.\frac{2\bar{c}''}{rB_0^2}\right]\frac{A_0^2}
{\left(\frac{b}{B_0} -\frac{2\bar{c}}{r}\right)}.
\end{eqnarray}
For the sake of instability region, we assume that all the
functions involved in the above equation are such that $\alpha$
remains positive. Consequently, the solution of Eq.(\ref{66})
becomes
\begin{equation}\label{68}
T(t)=-e^{\sqrt{\alpha}t}.
\end{equation}
Here we assume that the system starts collapsing at $t=-\infty$ such
that $T(-\infty)=0$, keeping it in static position. Afterwards, it
goes on collapsing with the increase of $t$.

Applying static and non-static perturbation scheme respectively on
dynamical equations (\ref{28}) and (\ref{29}), we have
\begin{eqnarray}\label{59}
p_{r0}'+(\rho_0+p_{r0})\frac{A_0'}{A_0}+\frac{2}{r}(p_{r0}-p_{\perp0})&=&0,\\\label{60}
\frac{1}{A_0}\left[\dot{\bar{\rho}}+(\rho_0+p_{r0})\dot{T}\frac{b}{B_0}
+2(\rho_0+p_{\perp0})\dot{T}\frac{\bar{c}}{r}\right]&=&0,\\\nonumber
\frac{1}{B_0}\left[\bar{p_r}'+(\rho_0+p_{r0})T\left(\frac{a}{A_0}\right)'
+({\bar{\rho}}+\bar{p_r})\frac{A_0'}{A_0}\right.&&\\\label{61}
+\left.2(p_{r0}-p_{\perp0})T\left(\frac{\bar{c}}{r}\right)'
+\frac{2}{r}(\bar{p_{r}}-\bar{p}_{\perp})\right]&=&0.
\end{eqnarray}
Integrating Eq.(\ref{60}) with respect to time, it follows that
\begin{equation}\label{62}
\bar{\rho}=-\left[(\rho_0+p_{r0})\frac{b}{B_0}
+2(\rho_0+p_{\perp0})\frac{\bar{c}}{r}\right]T.
\end{equation}
Perturbation on Eq.(\ref{20}) yields
\begin{eqnarray}\label{63}
m_0&=&\frac{r}{2}\left(1-\frac{1}{B_0^2}\right),\\\label{64}
\bar{m}&=&-\frac{T}{B_0^2}\left[r\left(\bar{c}'-\frac{b}{B_0}\right)
+(1-B_0^2)\frac{\bar{c}}{2}\right].
\end{eqnarray}
In order to relate $\bar{\rho}$ and $\bar{p}_r$ for the static
spherically symmetric configuration, we may assume an equation of
state of Harrison-Wheeler type as follows (Chan et al. 239;
Wheeler et al. 1965)
\begin{equation}\label{69}
\bar{p}_r=\Gamma\frac{p_{r0}}{\rho_0+p_{r0}}\bar{\rho},
\end{equation}
where $\Gamma$ is the adiabatic index. It measures the variation
of pressure for a given variation of density. We take it constant
throughout the fluid evolution.

\section{Expansionfree Newtonian and Post Newtonian Regimes}

Here we assume expansionfree ($\Theta=0$) evolution of anisotropic
fluid and develop dynamical equations. Since the expansion scalar
describes the rate of change of small volumes of the fluid, so the
expansionfree fluid evolution of spherically symmetric distribution
is consistent with the formation of a vacuum cavity within the
fluid. The Minkowski spacetime is supposed to present inside the
cavity. Using Eq.(\ref{48'}) in (\ref{15}), we have
\begin{equation}\setcounter{equation}{1}\label{57}
\bar{\Theta}=\frac{\dot{T}}{A_0}\left(\frac{b}{B_0}+\frac{2\bar{c}}{r}\right).
\end{equation}
The expansionfree condition ($\bar{\Theta}=0$) implies that
\begin{equation}\label{79}
\frac{b}{B_0}=-2\frac{\bar{c}}{r}.
\end{equation}
Before implementing the above result on the last section, let us
identify the terms differentiating Newtonian (N), post Newtonian
(pN) and post post Newtonian (ppN) regimes. These terms will be
considered to develop the dynamical equations which help to
understand the instability conditions of the expansionfree fluid
evolution. For N approximation, we assume
\begin{equation}\label{70}
\rho_0\gg p_{r0},\quad\rho_0\gg p_{\perp0}.
\end{equation}

For the metric coefficients expanded up to pN approximation, we
take
\begin{equation}\label{71}
A_0=1-\frac{Gm_0}{c^2},\quad B_0=1+\frac{Gm_0}{c^2},
\end{equation}
where $G$ is the gravitational constant and $c$ is the speed of
light. Adding Eqs.(\ref{16s}) and (\ref{18s}) and using
Eq.(\ref{63}), it follows that
\begin{equation}\label{72}
\frac{A_0'}{A_0}=\frac{\kappa r^3(\rho_0+p_{r0})(1+2\delta
R_0)+2\delta R_0''r^2(r-2m_0)+2m_0(1+2\delta R_0+\delta
R_0'r)}{2r(r-2m_0)(1+2\delta R_0+\delta R_0'r)},
\end{equation}
Inserting this equation in (\ref{59}), we have first dynamical
equation in relativistic units as follows
\begin{eqnarray}\nonumber
p_{r0}'&=&\frac{-(\rho_0+p_{r0})}{2r(r-2m_0)(1+2\delta R_0)+\delta
R_0'r)}[\kappa r^3(\rho_0+p_{r0})(1+2\delta
R_0)\\\nonumber&+&2\delta R_0''r^2(r-2m_0)+2m_0(1+2\delta
R_0+\delta R_0'r)]\\\label{73}&+&\frac{2}{r}(p_{\perp0}-p_{r0}).
\end{eqnarray}
In view of dimensional analysis, this equation can be written in
c.g.s. units as follows
\begin{eqnarray}\nonumber
p_{r0}'&=&\frac{-(\rho_0+c^{-2}p_{r0})}{2rc^{-2}(r-2Gc^{-2}m_0)(1+2\delta
R_0+\delta R_0'r)}[G\kappa r^3(\rho_0+c^{-2}p_{r0})(1+2\delta
R_0)\\\nonumber &+&2\delta
R_0''r^2(r-2Gc^{-2}m_0)+2Gc^{-2}m_0(1+2\delta R_0+\delta
R_0'r)]\\\label{74} &+&\frac{2}{r}(p_{\perp0}-p_{r0}).
\end{eqnarray}
Expanding up to $c^{-4}$ order, we get
\begin{eqnarray}\nonumber
p_{r0}'&=&\frac{2}{r}(p_{\perp0}-p_{r0})-\frac{\rho_0}{r^3}(r+Gm_0)[G\kappa
r^3\rho_0(1+2\delta R_0)+2\delta R_0''r^3](1-2\delta
R_0\\\nonumber &-&\delta
R_0'r)-\frac{G}{c^{2}r^3}[\rho_0(r+Gm_0)\{p_{r0}\kappa
r^3(1+2\delta R_0)-4\delta R_0''m_0r^2\\\nonumber
&+&2m_0(1+2\delta R_0+\delta R_0'r)\}+2\rho_0Gm_0^2r^2\{G\kappa
\rho_0(1+2\delta R_0)+2\delta R_0''\}\\\nonumber
&+&r^3(r+Gm_0)p_{r0}\{\kappa \rho_0(1+2\delta R_0)\}](1-2\delta
R_0-\delta R_0'r)\\\nonumber
&-&\frac{G}{c^{4}r^4}[rp_{r0}(r+Gm_0)\{p_{r0}\kappa r^3(1+2\delta
R_0)-4\delta R_0''m_0r^2+2m_0(1+2\delta R_0\\\nonumber &+&\delta
R_0'r)\}+p_{r0}Gm_0^2r^3\{G\kappa \rho_0(1+2\delta R_0)+2\delta
R_0''\}+2\rho_{0}G^2m_0^2\{\kappa r^3 p_{r0}\\\nonumber
&\times&(1+2\delta R_0)-4\delta R_0''m_0r^2+2m_0(1+2\delta
R_0+\delta R_0'r)\}](1-2\delta R_0-\delta R_0'r).\\\label{75}
\end{eqnarray}
It is worth mentioning here that the order of $c$ differentiates
the terms belonging to N, pN and ppN regimes, i.e.,
\begin{eqnarray}\label{76}
&&\textmd{terms of order}~ c^0~ \textmd{correspond to
N-approximation,}
\\\label{77} &&\textmd{terms of order} ~c^{-2}~ \textmd{correspond to
pN-approximation,}\\\label{78} &&\textmd{terms of order}~ c^{-4} ~
\textmd{correspond to ppN-approximation}.
\end{eqnarray}
Simplification of Eq.(\ref{75}) yields the following table
\begin{center}
{\bf {\small Table 1. N, pN and ppN Terms}} \vspace{0.25cm}
\vspace{0.25cm}
\begin{tabular}{|l|l|}
\hline {\bf N terms} &$p_{r0}$, $p_{\perp 0}$,
$\rho_{0}^2(1+2\delta R_0)$, $m_0\rho_{0}^2(1+2\delta R_0)$,
$\rho_0\delta R_0''$,
$m_0\rho_0\delta R_0''$ \\
\hline {\bf pN terms} &$p_{r0}$$\rho_{0}(1+2\delta R_0)$,
$p_{r0}\rho_{0}m_0(1+2\delta R_0)$, $m_0^2\rho_{0}^2(1+2\delta
R_0)$, \\& $m_0^2\delta R_0''$, $m_0^2\rho_0\delta R_0''$,
$p_{r0}\delta R_0''$, $m_0^2(1+2\delta
R_0+\delta R_0')$ \\&$m_0\rho_0(1+2\delta R_0+\delta R_0')$\\
\hline {\bf ppN terms} &$p_{r0}^2(1+2\delta R_0)$,
$p_{r0}\rho_{0}m_0^2(1+2\delta R_0)$, $m_0p_{r0}^2(1+2\delta
R_0)$, \\& $m_0^2p_{r0}\delta R_0''$, $\rho_0m_0^3\delta R_0''$,
$m_0p_{r0}\delta R_0''$, $p_{r0}m_0(1+2\delta
R_0+\delta R_0')$ \\&$m_0^2p_{r0}(1+2\delta R_0+\delta R_0')$,
$m_0^3\rho_{0}(1+2\delta R_0+\delta R_0')$\\
\hline
\end{tabular}
\end{center}

Using the expansionfree condition (\ref{79}), Eq.(\ref{62}) becomes
\begin{equation}\label{82}
\bar{\rho}=2(p_{r0}-p_{\perp0})\frac{T\bar{c}}{r}.
\end{equation}
This shows that the perturbed energy density depends on the static
pressure anisotropy, hence supporting the expansionfree condition.
Inserting Eq.(\ref{82}) in (\ref{69}), we have
\begin{equation}\label{83}
\bar{p}_r=2\Gamma\frac{p_{r0}}{\rho_0+p_{r0}}(p_{r0}-p_{\perp0})\frac{T\bar{c}}{r}~.
\end{equation}
From Eq.(\ref{59}), we can write
\begin{eqnarray}\label{84}
\frac{A_0'}{A_0}=\frac{-1}{\rho_0+p_{r0}}\left[p_{r0}'+\frac{2}{r}(p_{r0}-p_{\perp0})\right].
\end{eqnarray}
Also, from Eq.(\ref{63}), we obtain
\begin{equation}\label{85}
\frac{B_0'}{B_0}=\frac{-m_0}{r(r-2m_0)}.
\end{equation}
Notice that the expressions of $\bar{p}_r$ and
$\bar{\rho}\frac{A_0'}{A_0}$ in Eqs.(\ref{82})-(\ref{84}) are of ppN
order approximation. In order to discuss the instability conditions
up to pN order, we neglect these quantities.

Substituting the value of $\bar{p}_{\perp}$ from Eq.(\ref{19p}) in
dynamical equation (\ref{61}) and using Eqs.(\ref{68}) and
(\ref{79}), it follows that
\begin{eqnarray}\nonumber
&&\kappa(\rho_0+p_{r0})r\left(\frac{a}{A_0}\right)'+2\kappa
r(p_{r0}-p_{\perp0})\left(\frac{c}{r}\right)'-8\kappa
p_{\perp0}\frac{\bar{c}}{r}-\frac{2}{B_0^2}\left[\frac{1}{r}\left(\frac{a}{A_0}\right)'
\right.\\\nonumber
&+&\left.\left(\frac{\bar{c}}{r}\right)'\left(\frac{A_0'}{A_0}+\frac{B_0'}{B_0}
-\frac{2}{r}\right)
+\left(\frac{\bar{c}}{r}\right)''-\frac{2c}{r^3}(1+3B_0^2)+\frac{a}{r}\frac{A_0'}{A_0}+
\frac{8\bar{c}}{r^2}\left(\frac{A_0'}{A_0}+\frac{B_0'}{B_0}\right)\right]\\\nonumber
&+&e-\frac{2R_0(1+\delta R_0)}{1+2\delta
R_0}\left(\frac{2\bar{c}}{r}-\frac{\delta e}{1+2\delta
R_0}\right)+\frac{4\delta R_0'}{B_0^2(1+2\delta
R_0)}\left\{\left(\frac{a}{A_0}\right)'+3\left(\frac{\bar{c}}{r}\right)'
\right\}\\\nonumber &+&\frac{4\delta }{B_0^2(1+2\delta
R_0)}\left(e''-\frac{2\delta eR_0''}{1+2\delta R_0}\right)-
\frac{4\delta}{B_0^2}\left(\frac{e}{1+2\delta
R_0}\right)'\left(\frac{A_0'}{A_0}-\frac{B_0'}{B_0}+\frac{1}{r}\right)\\\label{86}
&+&\frac{2\alpha}{A_0^2}\left(\frac{\bar{c}}{r}+\frac{2e\delta}{1+2\delta
R_0} \right)=0.
\end{eqnarray}
This shows that the general dependence of radial function affects
dynamics of collapsing fluid. For the sake of simplicity, we assume
\begin{eqnarray}\label{a}
a(r)=a_0+a_1r,\quad \bar{c}(r)=c_0+c_1r,\quad e(r)=e_0+e_1r,
\end{eqnarray}
where the quantities with subscript "0" and "1" are arbitrary
constants. Using these values in Eq.(\ref{86}), it follows that
\begin{eqnarray}\nonumber
&&\kappa(\rho_0+p_{r0})r\frac{a_1}{A_0}- \kappa r
(\rho_0+p_{r0})(a_0+a_1r)\frac{A_0'}{A_0^2}-2\kappa
(p_{r0}+3p_{\perp0})\frac{c_0}{r}\\\nonumber&-&8\kappa
p_{\perp0}c_1-\frac{2}{B_0^2}\left[\frac{a_1}{A_0r}-\left(\frac{a_0}{r}+a_1\right)
(A_0-1)\frac{A_0'}{A_0^2}
-\frac{c_0}{r^2}\left(\frac{A_0'}{A_0}+\frac{B_0'}{B_0}-\frac{2}{r}\right)\right.\\\nonumber
&+&\left.\frac{4c_0}{r^3}+\frac{2c_1}{r^2}+
\frac{6}{r^3}(c_0+c_1r)B_0^2+\frac{8}{r^2}(c_0+c_1r)
\left(\frac{A_0'}{A_0}+\frac{B_0'}{B_0}\right)\right]+e_0+e_1r\\\nonumber
&+&\frac{2R_0(1+\delta R_0)}{1+2\delta
R_0}\left[\frac{2(c_0+c_1r)}{r}+\frac{\delta(e_0+e_1r)}{1+2\delta
R_0}\right]+\frac{4\delta R_0'}{B_0^2(1+2\delta
R_0)}\\\nonumber&\times&\left[\frac{a_1}{A_0}-(a_0+a_1r)\frac{A_0'}{A_0^2}
-\frac{3c_0}{r^2}\right]-\frac{8\delta^2
R_0''(e_0+e_1r)}{B_0^2(1+2\delta R_0)^2}-
\frac{4\delta}{B_0^2}\left(\frac{e_0+e_1r}{1+2\delta
R_0}\right)'\\\label{86'}&\times&\left(\frac{A_0'}{A_0}-\frac{B_0'}{B_0}+\frac{1}{r}\right)
+\frac{2\alpha}{A_0^2}\left[\frac{c_0+c_1r}{r}+\frac{2(e_0+e_1r)\delta}{1+2\delta
R_0} \right]=0.
\end{eqnarray}
Inserting Eqs.(\ref{71}), (\ref{84}) and (\ref{85}) up to pN order
(with $c=G=1$) in the above equation, we obtain
\begin{eqnarray}\nonumber
&&\kappa(\rho_0+p_{r0})a_1r+\kappa m_0 (\rho_0+p_{r0})+\kappa r
(a_0+a_1r)\left(1+\frac{m_0}{r}\right)\\\nonumber&\times&
\left[p_{r0}'+\frac{2}{r}(p_{r0}-p_{\perp0})\right]-2\kappa
(p_{r0}+3p_{\perp0})\frac{c_0}{r}-8\kappa
p_{\perp0}c_1-2\left(1-\frac{m_0}{r}\right)\\\nonumber&\times&
\left[a_1+\frac{a_1m_0}{r^2}-\frac{1}{\rho_0}\left\{p_{r0}'+\frac{2}{r}(p_{r0}
-p_{\perp0})\right\}\left(1-\frac{p_{r0}}{\rho_0}\right)
\left(1+\frac{m_0}{r}\right)\left(\frac{a_0}{r}\right.\right.
\\\nonumber&+&\left.\left.a_1\right)+
\frac{10c_0}{r^3}+\frac{8c_1}{r^2}+\frac{12m_0}{r^4}(c_0+c_1r)+
\frac{c_0}{r^2\rho_0}\left\{p_{r0}'+\frac{2}{r}(p_{r0}-p_{\perp0})\right\}
\right.\\\nonumber&\times&\left.
\left(1-\frac{p_{r0}}{\rho_0}\right)+
\frac{c_0m_0}{r^4}\left(1+\frac{2m_0}{r}\right)-\frac{2c_0}{r^3}
-\left(\frac{a_0}{r}+a_1\right)\frac{1}{\rho_0}\left\{p_{r0}'+\frac{2}{r}(p_{r0}
\right.\right.\\\nonumber&-&\left.\left.
p_{\perp0})\right\}\left(1-\frac{p_{r0}}{\rho_0}\right)
-\frac{8(c_0+c_1r)}{r^2\rho_0}\left\{p_{r0}'+\frac{2}{r}(p_{r0}-p_{\perp0})\right\}
\left(1-\frac{p_{r0}}{\rho_0}\right)\right.
\end{eqnarray}
\begin{eqnarray}\nonumber&+&\left.
\frac{8(c_0+c_1r)m_0}{r^4}\left(1+\frac{2m_0}{r}\right)
\right]+e_0+e_1r+\frac{2R_0(1+\delta R_0)}{1+2\delta
R_0}\left[\frac{2(c_0+c_1r)}{r}\right.\\\nonumber
&+&\left.\frac{\delta(e_0+e_1r)}{1+2\delta
R_0}\right]+\frac{4\delta R_0'}{(1+2\delta
R_0)}\left(1-\frac{2m_0}{r}\right)\left[a_1+\frac{a_1m_0}{r}-\frac{3c_0}{r^2}\right.
\\\nonumber&+&\left.(a_0+a_1r)\left(1+\frac{m_0}{r}\right)\frac{1}{\rho_0}
\left\{p_{r0}'+\frac{2}{r}(p_{r0}-p_{\perp0})\right\}
\left(1-\frac{p_{r0}}{\rho_0}\right)\right]\\\nonumber&+&\frac{8\delta^2
R_0''(e_0+e_1r)}{(1+2\delta R_0)^2}\left(1-\frac{2m_0}{r}\right)
+4\delta\left(\frac{e_0+e_1r}{1+2\delta
R_0}\right)'\left(1-\frac{2m_0}{r}\right)\\\nonumber&\times&
\left[\frac{-1}{\rho_0}
\left\{p_{r0}'+\frac{2}{r}(p_{r0}-p_{\perp0})\right\}
\left(1-\frac{p_{r0}}{\rho_0}\right)+\frac{m_0}{r^2}\left(1+\frac{2m_0}{r}\right)+\frac{1}{r}
\right]\\\label{86''}&+&2\alpha\left(1+\frac{2m_0}{r}\right)\left[\frac{c_0+c_1r}{r}
+\frac{2(e_0+e_1r)\delta}{1+2\delta R_0} \right]=0.
\end{eqnarray}

In view of \textbf{Table 1} and neglecting terms with
$p_{r0}/\rho_0$ being of ppN order, Eq.(\ref{86''}) reduces to the
following equation in the Newtonian regime
\begin{eqnarray}\nonumber
&&\kappa(r+m_0)\rho_0+\kappa
p_{r0}(3a_1r+2a_0-\frac{2c_0}{r})+2\kappa
p_{\perp0}(a_0+a_1r+\frac{3c_0}{r}-8c_1)\\\nonumber &+&\kappa
r(a_0+a_1r)|p_{r0}'|+e_0+e_1r-2a_1-\frac{20c_0}{r^3}-\frac{16c_1}{r^2}
+\frac{2}{r}\\\nonumber
&+&\frac{m_0}{r^2}\left[a_1r+\frac{2c_0}{r}(4r+5)+\frac{8c_1}{r}(r+1)\right]
\\\nonumber &+& \frac{2R_0(1+\delta R_0)}{1+2\delta
R_0}\left[\frac{2(c_0+c_1r)}{r}+\frac{\delta(e_0+e_1r)}{1+2\delta
R_0}\right]\\\nonumber &+&\frac{4\delta R_0'}{1+2\delta
R_0}\left[a_1-\frac{3c_0}{r^2}+\frac{m_0}{r^2}(2-a_1r)\right]
+\frac{4\delta}{r}\left(\frac{e_0+e_1r}{1+2\delta
R_0}\right)'\\\label{95}&+&2\alpha\left(1+\frac{2m_0}{r}\right)\left[\frac{c_0+c_1r}{r}
+\frac{2(e_0+e_1r)\delta}{1+2\delta R_0} \right]=0.
\end{eqnarray}
In general, the instability range depends upon the index $\Gamma$ as
it measures the compressibility of the fluid. However, the above
equation is independent of $\Gamma$ which shows that instability
region totally depends upon the pressure anisotropy, energy density,
chosen $f(R)$ model and arbitrary constants. Notice that
independence of $\Gamma$ factor indicates that under expansionfree
condition, fluid evolves without being compressed. In this way, the
given $f(R)$ model shows the consistency of the physical results
with expansionfree condition.

In order to satisfy the instability conditions of expansionfree
fluids, we need to keep all the terms positive in Eq.(\ref{95}).
Here we assume that all the arbitrary constants and dynamical
quantities are positive whereas $p_{r0}'<0$ shows that pressure
decreases during collapsing process. To keep all the terms in
dynamical equations positive in the Newtonian regime, we need to
satisfy the following constraints arising from different terms
\begin{eqnarray}\label{in1}
&&\frac{3c_0}{r^2}<a_1<\frac{2}{r},\\\label{in2}
&&a_0+a_1r+\frac{3c_0}{r}>8c_1,\\\label{in3}
&&2a_0+3a_1r>\frac{2c_0}{r},\\\label{in4}
&&0<a_1r+\frac{10c_0}{r^2}+\frac{8c_1}{r}<1.
\end{eqnarray}
Thus the system would be unstable in N-approximation as long as
the above inequalities are satisfied. The dynamical equation
(\ref{86''}) for pN regime becomes
\begin{eqnarray}\nonumber
&&3\kappa(\rho_0+p_{r0})a_1r+\kappa m_0 (\rho_0+p_{r0})+\kappa r
(a_0+a_1r)\left(1+\frac{m_0}{r}\right)\\\nonumber&\times&
\left[p_{r0}'+\frac{2}{r}(p_{r0}-p_{\perp0})\right]-2\kappa
(p_{r0}-3p_{\perp0})\frac{c_0}{r}-8\kappa
p_{\perp0}c_1-2\left(1-\frac{m_0}{r}\right)
\end{eqnarray}
\begin{eqnarray}\nonumber
&\times&\left[a_1+\frac{a_1m_0}{r}+\frac{10c_0}{r^3}+\frac{8c_1}{r^2}
+\frac{c_0}{r^2}\left\{\frac{m_0}{r^2}\left(1+\frac{2m_0}{r}\right)-\frac{2}{r}\right\}
+\frac{8m_0(c_0+c_1r)}{r^4}\right.\\\nonumber&\times&\left.\left(1+\frac{2m_0}{r}\right)\right]
+e_0+e_1r+\frac{2R_0(1+\delta R_0)}{1+2\delta
R_0}\left[\frac{2(c_0+c_1r)}{r}+\frac{\delta(e_0+e_1r)}{1+2\delta
R_0}\right]\\\nonumber&+&\frac{4\delta R_0'}{(1+2\delta
R_0)}\left(1-\frac{2m_0}{r}\right)\left[a_1+\frac{a_1m_0}{r}-\frac{3c_0}{r^2}\right]
+\frac{8\delta^2R_0''(e_0+e_1r)}{(1+2\delta
R_0)^2}\\\nonumber&\times&\left(1-\frac{2m_0}{r}\right)
+4\delta\left(\frac{e_0+e_1r}{1+2\delta
R_0}\right)'\left(1-\frac{2m_0}{r}\right)\left(\frac{m_0}{r^2}
+\frac{1}{r}\right)\\\label{97}&+&2\alpha\left(1+\frac{2m_0}{r}\right)\left[\frac{c_0+c_1r}{r}
+\frac{2(e_0+e_1r)\delta}{1+2\delta R_0} \right]=0.
\end{eqnarray}
It is remarked that only relativistic effects are taken into account
at pN regime, however, the dependence of instability condition
remains the same. Also, the index $\Gamma$ does not involve, so
instability conditions depend on the same parameters and constants
as that in the N-approximation.

\section{Summary}

This paper is devoted to investigate the dynamical instability
conditions of the expansionfree fluid evolution for a particular
model ($f(R)=R+\delta R^2$) in $f(R)$ gravity. To make consistency
with the physical application of the vanishing $\Theta$, we have
considered the locally anisotropic fluid with inhomogeneous energy
density. Perturbation analysis is used in N and pN approximations
and terms differentiated in both the regimes depending upon the
order of $c$.

Expansionfree models may help to study the formation of voids.
Voids are the spongelike structures and occupying 40\%-50\% volume
of the universe. Observations suggest very different sizes of the
voids, i.e., mini-voids (Tikhonov et al. 2006) to super-voids
(Rudnick et al. 2007). As concerned to the shapes of the voids,
they are neither empty nor spherical. However, for the sake of
investigations, they are usually described as vacuum spherical
cavities surrounding by the fluid. The assumption of spherically
symmetric spacetime outside the cavity is justified for cavities
with sizes of the order of 20 Mpc or smaller, as the observed
universe would be inhomogeneous on scale less than 150-300 Mpc.

The adiabatic index $\Gamma$ measures the variation of pressure,
its value defines the range of instability. For example, for a
Newtonian perfect fluid, the system is unstable for $\Gamma<4/3$.
We have found that like GR (Herrera et al. arXiv/1010.1518), our
results are also independent of the $\Gamma$ factor. This shows
the consistency of expansionfree condition with $f(R)$ gravity as
this requires that fluid should evolve without compressibility.
Moreover, the instability range depends upon the radial anisotropy
of pressure, energy density and some constraints on the constants
arising from the positivity of the dynamical equation. Equation
(\ref{95}) holds the instability requirement as long as the
inequalities (\ref{in1})-(\ref{in4}) are satisfied.

It is worthwhile to mention here that the chosen $f(R)$ model is the
only model admitting Schwarzschild solution. We have used the
perturbed Ricci scalar to get the time dependent part of the
perturbed metric coefficients. In this way, dynamics of the
gravitational collapse comprises the scalar curvature in its
evolution.

\vspace{0.5cm}

{\bf Acknowledgment}

\vspace{0.25cm}

We would like to thank the Higher Education Commission, Islamabad,
Pakistan for its financial support through the {\it Indigenous
Ph.D. 5000 Fellowship Program Batch-III}.

\section*{References}

Carroll, S.M., et al.: New J. Phys. \textbf{8}, 323(2006)\\
Carloni, S., Dunsby, P.K.S. and Troisi, A.: Phys. Rev.
\textbf{D77}, 024024(2008)\\
Chandrasekhar, S.: Astrophys. J. \textbf{140}, 417(1964)\\
Chiba, T.: Phys. Lett. \textbf{B575}, 1(2003)\\Chan, R., et al.:
Mon. Not. R. Astron. Soc. \textbf{239}, 91(1989)\\
Chan, R., Herrera, L. and Santos, N.O.: Mon. Not. R. Astron. Soc.
\textbf{267}, 637(1994)\\
Chan, R., Herrera, L. and Santos, N.O.: Mon. Not. R. Astron. Soc.
\textbf{265}, 533(1993)\\
Dolgov, A.D. and Kawasaki, M.: Phys. Lett. \textbf{B573}, 1(2003)\\
de la Cruz-Dombrize, A. and Dobado, A.: Phys. Rev. \textbf{D74},
087501(2006)\\
de la Cruz-Dombriz, A., Dobado, A. and Maroto, A.L.: Phys.Rev. \textbf{D80}, 124011(2009)\\
de la Cruz-Dombriz, A. Dobado, A. and Maroto, A.L.: Contribution
to the Proceedings of Spanish Relativity Meeting, Spain (2009)\\
Herrera, L. and Santos, N.O.: Phys. Rev. {\bf D70}, 084004(2004)\\
Herrera, L., Santos, N.O. and Wang, A.: Phys. Rev. \textbf{D78},
080426(2008). Herrera, L., Santos, N.O. and Le Denmat,
G.: Mon. Not. R. Astron. Soc. \textbf{237}, 257(1989)\\
Herrera, L., Santos, N.O. and Le Denmat, G.: \textit{Dynamical
Instability of Expansion-free Fluids}; arXiv/1010.1518\\
Hwang, J.C. and Noh. H.: Phys. Rev. \textbf{D54}, 1460(2006)\\
Joshi, P., Dadhich, N. and Maartens, R.: Phys. Rev.
\textbf{D65}, 101501(2002)\\
Liddle, A.R. and Wands, D.: Mon. Not. Roy. Astron. Soc.
\textbf{253}, 637(1991)\\
Misner, C.W. and Sharp, D.: Phys. Rev. \textbf{136}, B571(1964)\\
Mignemi, S. and Wiltshire, D.L.: Phys. Rev.
\textbf{D46}, 1475(1992)\\
Rudnick, L., Brown, S. and Williams, L.R.: Astrophys. J.
\textbf{671}, 40(2007)\\
Sharif, M. and Kausar, H.R.: Int. J. Mod. Phys. \textbf{D20},
2239(2011)\\
Sharif, M. and Kausar, H.R.: Mod. Phys. Lett. \textbf{A25}, 3299(2010)\\
Sharif, M. and Kausar, H.R.: Astrophys. Space Sci. \textbf{331},
281(2011)\\
Sharif, M. and Kausar, H.R.: Astrophys. Space Sci. \textbf{332}, 463(2011)\\
Sharif, M. and Kausar, H.R.: J. Phys. Soc. Jpn. \textbf{80}, 044004(2011)\\
Sharif, M. and Kausar, H.R.: Phys. Lett. \textbf{B697}, 01(2011)\\
Sharif, M. and Kausar, H.R.: JCAP \textbf{07}, 022(2011)\\
Skripkin, V.A.: Soviet Physics-Doklady \textbf{135}, 1183(1960)\\
Sotiriou, T.P. and Faraoni, V.: Rev. Mod. Phys. \textbf{82},
451(2010)\\
Tikhonov, A.V. and Karachentsev, I.D.: Astrophys. J. \textbf{653},
969(2006)\\ Tsujikawa, S., Uddin, K. and Tavakol, R.: Phys. Rev.
\textbf{D77}, 043007(2008)\\ Mazharimousavi, S.H. and Halilsoy, M.:
Phys. Rev. \textbf{D84}, 064032(2011)\\ Moon, T., Myung,
Y.S. and Son, E.J.: \emph{$f(R)$ Black Holes}; arXiv:1101.1153\\
Whitt, B.: Phys. Lett. \textbf{B145}, 176(1984)\\
Wheeler, J.A., et al.: Gravitation and Gravitational Collapse.
University of Chicago Press (1965)
\end{document}